\documentstyle[epsf]{mn}

\newif\ifAMStwofonts

\ifoldfss
  \ifCUPmtlplainloaded \else
    \NewTextAlphabet{textbfit} {cmbxti10} {}
    \NewTextAlphabet{textbfss} {cmssbx10} {}
    \NewMathAlphabet{mathbfit} {cmbxti10} {} 
    \NewMathAlphabet{mathbfss} {cmssbx10} {} 
  \fi
  \ifAMStwofonts
    \ifCUPmtlplainloaded \else
      \NewSymbolFont{upmath} {eurm10}
      \NewSymbolFont{AMSa} {msam10}
      \NewMathSymbol{\upi}     {0}{upmath}{19}
      \NewMathSymbol{\umu}     {0}{upmath}{16}
      \NewMathSymbol{\upartial}{0}{upmath}{40}
      \NewMathSymbol{\leqslant}{3}{AMSa}{36}
      \NewMathSymbol{\geqslant}{3}{AMSa}{3E}

    \fi
  \fi
\fi 

\ifnfssone
  \newmathalphabet{\mathit}
  \addtoversion{normal}{\mathit}{cmr}{m}{it}
  \addtoversion{bold}{\mathit}{cmr}{bx}{it}
  \newmathalphabet{\mathbfit} 
  \addtoversion{normal}{\mathbfit}{cmr}{bx}{it}
  \addtoversion{bold}{\mathbfit}{cmr}{bx}{it}
  \newmathalphabet{\mathbfss} 
  \addtoversion{normal}{\mathbfss}{cmss}{bx}{n}
  \addtoversion{bold}{\mathbfss}{cmss}{bx}{n}
  \ifAMStwofonts
    \ifCUPmtlplainloaded \else
      \UseAMStwoboldmath
      \makeatletter
      \new@mathgroup\upmath@group
      \define@mathgroup\mv@normal\upmath@group{eur}{m}{n}
      \define@mathgroup\mv@bold\upmath@group{eur}{b}{n}
      \edef\UPM{\hexnumber\upmath@group}
      \new@mathgroup\amsa@group
      \define@mathgroup\mv@normal\amsa@group{msa}{m}{n}
      \define@mathgroup\mv@bold\amsa@group{msa}{m}{n}
      \edef\AMSa{\hexnumber\amsa@group}
      \makeatother
      \mathchardef\upi="0\UPM19
      \mathchardef\umu="0\UPM16
      \mathchardef\upartial="0\UPM40
      \mathchardef\leqslant="3\AMSa36
      \mathchardef\geqslant="3\AMSa3E
    \fi
  \fi
\fi 

\ifnfsstwo
  \DeclareMathAlphabet{\mathbfit}{OT1}{cmr}{bx}{it}
  \SetMathAlphabet\mathbfit{bold}{OT1}{cmr}{bx}{it}
  \DeclareMathAlphabet{\mathbfss}{OT1}{cmss}{bx}{n}
  \SetMathAlphabet\mathbfss{bold}{OT1}{cmss}{bx}{n}
  \ifAMStwofonts
    \ifCUPmtlplainloaded \else
      \DeclareSymbolFont{UPM}{U}{eur}{m}{n}
      \SetSymbolFont{UPM}{bold}{U}{eur}{b}{n}
      \DeclareSymbolFont{AMSa}{U}{msa}{m}{n}
      \DeclareMathSymbol{\upi}{0}{UPM}{"19}
      \DeclareMathSymbol{\umu}{0}{UPM}{"16}
      \DeclareMathSymbol{\upartial}{0}{UPM}{"40}
      \DeclareMathSymbol{\leqslant}{3}{AMSa}{"36}
      \DeclareMathSymbol{\geqslant}{3}{AMSa}{"3E}
    \fi
  \fi
\fi 

\ifCUPmtlplainloaded \else
  \ifAMStwofonts \else 
    \def\upi{\pi}
    \def\umu{\mu}
    \def\upartial{\partial}
  \fi
\fi

\title[Separation of foregrounds for CMB mapping]
{Separation of instrumental and astrophysical foregrounds for 
mapping CMB anisotropies}
\author[J. Delabrouille, G. Patanchon \& E. Audit]
       {J. Delabrouille$^1$, G. Patanchon$^1$ \& E. Audit$^2$ \\
$^1$PCC --- Coll{\`e}ge de France, 11, place Marcelin Berthelot, 
F-75231 Paris, France\\
$^2$CEA/DSM/DAPNIA Service d'Astrophysique, CEA/Saclay 91191 
Gif-sur-Yvette Cedex, France}
\date{Accepted 2001 January 00.
      Received 2001 January 00;
      in original form 2001 January 00}

\pagerange{\pageref{firstpage}--\pageref{lastpage}}
\pubyear{2001}

\begin{document}

\maketitle

\label{firstpage}

\begin{abstract}
    
For the most sensitive present and future experiments dedicated to 
Cosmic Microwave Background (CMB) anisotropy observations, the 
identification and separation of signals coming from different sources 
is an important step in the data analysis.  This problem of the 
restitution of signals from the observation of their mixture is 
classically called ``component separation" in CMB mapping.  In this 
paper, we address the general problem of separating, for 
millimeter-wave 
sky mapping applications, components which include not only 
astrophysical emissions in two-dimensional maps, but also 
one-dimensional instrumental effects in the data streams.  We show 
that component separation methods can be adapted to separate 
simultaneously both astrophysical emissions and components coming from 
time-dependent foreground signals originating from the instrument 
itself.  Such general methods can be used for the optimal processing 
of low-redundancy observations where multi-channel observations are a 
precious tool to remove systematic effects, as may be the case for the 
Planck mission.

\end{abstract}


\section{Introduction}


Mapping and interpreting sky emissions in the millimeter and 
sub-millimeter range, and in particular Cosmic Microwave Background 
(CMB) temperature anisotropies, is one of the main objectives of 
present and upcoming observational effort in millimeter-wave 
astronomy.  This is, in particular, the objective of balloon and 
space--borne missions as Archeops \cite{archeops-inst}, Boomerang 
\cite{boomerang-inst}, Maxima \cite{Maxima}, TopHat \cite{Tophat}, MAP 
\cite{Map} and Planck \cite{lamarre00,bersanelli00}.  Among the scientific 
objectives of these missions, the precise measurement of primordial 
temperature and/or polarisation fluctuations of the CMB is one of the 
priorities.

The importance of measuring anisotropies of the Cosmic Microwave 
Background (CMB) to constrain cosmological models is now well 
established.  In the past ten years, tremendous theoretical activity, 
motivated by upcoming ambitious balloon-borne and space missions, 
demonstrated that measuring the properties of these temperature 
anisotropies will constrain drastically the seeds for structure 
formation as well as the cosmological parameters describing the matter 
content, the geometry, and the evolution of our Universe
\cite{hu-sugiyama96,jungman96}.  The accuracies required for precision 
tests of the cosmological models and measurements of the 
cosmological parameters (the Hubble constant $H_{0}$, the total matter
density $\Omega_{m}$, the cosmological constant $\Lambda$, \ldots), 
however, is such that it is necessary to disentangle in the data the 
contribution of several distinct astrophysical sources, all of which 
emit radiation in the frequency range used for CMB observations, 
i.e. the 10 GHz - 800 GHz range.


This problem, now traditionnally called ``component separation" by the 
CMB community, has been adressed already by a number of authors 
\cite{tegmark-efstathiou96,hjlb98,fb-rg99,bps99,baccigalupi00,snoussi01,maino01} 
using different assumptions and analysis techniques.  It is a 
particular application of the problem of ``source separation", also 
called in signal processing the ``cocktail party" problem, in which a 
number of source signals (timestreams, images\ldots) has to be 
recovered from the observation of their mixture in a number of 
captors.

So far however, only the separation of astrophysical components has 
been adressed using a component separation formalism. In addition, the 
separation has been implemented for ideal white noise on the observed 
maps. Real measurements however often suffer from instrumental 
effects, coloured noise, and involve the reprojection of 
one-dimensional data streams onto somewhat irregularly sampled two-dimensional 
maps, where some pixels may be observed several times and/or others not 
covered at all by the observations, depending on the resolution of the 
reconstructed map. In this paper, we investigate the possibility to 
extend the Wiener solution discussed by Tegmark \& Efstathiou 
\shortcite{tegmark-efstathiou96} and Bouchet \& Gispert
\shortcite{fb-rg99} to deal also with instrumental foregrounds. We
test the method in more realistic situations which include non-white detector 
noise and uncertainties induced by the reprojection of timelines.

The remaining of this paper is organised as follows: In section 
\ref{sec:comp-sep}, we introduce the general formalism of component 
separation and discuss the possibilty to include instrumental 
components and their specificities.  In section \ref{sec:sim}, we 
discuss the simulation of Planck observations containing the 
superposition of CMB and thermal dust emission, an instrumental 
effect, low-frequency detector noise, and white noise.  In section 
\ref{sec:imp-comp-sep}, we discuss the Wiener method implemented to 
separate these components in the simulated Planck data, and show the 
results we obtain.  In section \ref{sec:disc}, we discuss the results 
and the generality of the method.  We conclude in the last 
section.

\section{Component separation} \label{sec:comp-sep}

One of the key aspects of the component separation is the exploitation 
of the presumably known (at least to a large extent) dependence of the 
component emission on the frequency of observation $\nu$.  Although it 
is not always necessary that the frequency scalings be known {\it a 
priori} \cite{baccigalupi00,snoussi01,maino01}, separation techniques rely on 
the existence of such a frequency scaling.  In this framework, for 
multifrequency observations of the sky, the observed emission at 
frequency $\nu$ is the linear superposition of the emission of several 
components:
\begin{equation}
    I({\vec n},\nu) = \sum_{c} I_{c}({\vec n},\nu)
\end{equation}
where $I_{c}({\vec n},\nu)$ stands for the emission of component $c$ 
as a function of direction ${\vec n}$ and frequency $\nu$.  For each 
component, the emission can be well approximated as the product of a 
frequency spectrum $A_{c}(\nu)$ and a spatial template $\Delta({\vec 
n})$.
\begin{equation}
    I_{c}({\vec n},\nu) = A_{c}(\nu) \Delta_{c}({\vec n})
    \label{eq:spec-template}
\end{equation}
This approximation is very good for small portions of the sky.  
For larger maps (especially covering both low and high galactic 
latitudes), there can be variations of the emission law of the 
galactic components.

The decomposition in equation \ref{eq:spec-template} is not unique, as 
a multiplicative constant factor can be taken from $A$ into $\Delta$.  
For convenience, we adopt the convention that $\Delta_{c}$ is the 200 
GHz template of component $c$ in units of $\Delta T/T_{\rm CMB}$, that 
is:
\begin{equation}
    \Delta_{c}({\vec n}) = I_{c}({\vec n},200 {\, \rm GHz}) \frac{1}{T_{\rm CMB}}
    \left[ \frac{\partial B_{\nu}}{\partial T}(T_{\rm 
    CMB}, 200 {\, \rm GHz}) \right]^{-1}
\end{equation}
with $T_{\rm CMB}$ = 2.726 K, and $B_{\nu}$ is the Planck Blackbody spectrum.

For the present application, we will assume that the frequency spectra 
of astrophysical component emissions are known, which 
is a reasonable assumption in the 100-800 GHz frequency range, where 
the sky emission at moderate galactic latitude is dominated by CMB 
anisotropies and thermal galactic dust emission (see figure 
\ref{fig:foreground-spectra}).  The CMB anisotropy spectrum is the 
derivative of a blackbody with respect to temperature:
\begin{equation}
    A_{\rm CMB}(\nu) = \frac{\partial B_{\nu}}{\partial T}(T_{\rm 
    CMB}, \nu)
\end{equation}
Thermal dust emission is well fitted in this frequency range by a 
thermal spectrum with a $\nu^2$ emissivity and a temperature of 17.5 
K:
\begin{equation}
    A_{\rm dust}(\nu) = \left( \frac{\nu}{200 {\, \rm GHz}} \right) ^2 \, 
    \frac{B_{\nu}(17.5 {\rm K}, \nu)}{B_{\nu}(17.5 {\rm K}, 200 {\, \rm GHz})}
\end{equation}
This temperature of 17.5 K can change slightly (at the level of a few 
Kelvin) across the sky \cite{SFD98}, which is a motivation for a 
local treatment of component separation (as compared to full-sky 
methods).

In addition to the known frequency dependence, one can introduce 
priors on the spatial statistical properties of the foregrounds.  For 
Gaussian random fields, exploiting the prior knowledge of the 
autocorrelation leads to methods as the Wiener solution implemented 
and tested on simulations by Tegmark \& Efstathiou 
\shortcite{tegmark-efstathiou96} and Bouchet \& Gispert \shortcite{fb-rg99}.


In addition to diffuse galactic and extragalactic components, several 
populations of point-like sources are expected to contribute to the 
microwave sky emission.  Dedicated techniques are being developed for 
their extraction \cite{vielva01}.  At this point however, techniques 
developed to separate components in CMB experiments did not address 
the problem of non-astrophysical components, i.e. of signal 
fluctuations in detector readouts that are not due to astrophysical 
emission, fixed on the sky, but to the physical emission of other 
objects radiatively, or conductively, coupled to the detectors.

One exemple of such a ``foreground" is the thermal emission of a 
slightly emissive telescope which temperature fluctuates in time.  
Detectors experience a time-dependent additive fluctuation of their 
signals, which, reprojected on the sky according to each detector's 
specific scanning strategy in the process of recombining timelines 
into maps, generate fake position-dependent anisotropies on each 
detector's reconstructed sky map.

Another example, specific to bolometric detectors, is the signal 
generated by temperature fluctuations of the thermal bath required to 
cool the detectors to sub-kelvin temperatures.  Such fluctuations have 
been seen in detector timelines for the Archeops Balloon test flight 
\cite{archeops-inst}, and are expected to be present (at a much lower 
level) in Planck HFI data as well.  

Finally, one other interesting example is that of atmospheric emission 
for sub-orbital experiments, in which case time and space dependent 
fluctuations of the atmosphere temperature and composition in the line 
of sight generate fluctuations in the detectors timelines.

Because of the process of their generation, such foregrounds are seen 
by all detectors simultaneously.  However, as fields of views of all 
detectors of a CMB experiment as the current ones do not point towards 
the same point of the sky at the same time, they cannot be treated as 
another component on the sky.  It is a component which ``lives" in the 
time domain, and which is reprojected on the sky according to the 
scanning strategy, in a detector-dependent way.  As a consequence, a 
given pixel on the sky, being observed at different times by different 
detectors, is not polluted by the same instrumental foreground for 
different detectors.

Quite often, such instrumental foregrounds are essentially low 
frequency drifts, although one might imagine scan-synchronous
temperature fluctuations depending on the orientation of the 
experiment with respect to external sources of heat as the Sun, or 
line-spectra fluctuations induced by cyclic power dissipation or by 
vibrations within the experimental set-up (e.g. engines).

In the case of random slow drifts, a common way to handle this kind of 
systematics is simply to filter them out using a high-pass filter.  
This method has the merit of simplicity and efficiency, but reduces 
the sensitivity of the experiment to scales larger than $\omega_{\rm 
scan} \times f_{\rm filter}$, where $\omega_{\rm scan}$ is the 
scanning speed on the sky and $f_{\rm filter}$ the filter cutout 
frequency.  For experiments aiming at measuring anisotropies at all 
available scales, including the largest ones, more sophisticated 
destriping methods are implemented.  Such methods, whatever the exact 
algorithm used, essentially rely on redundancies (i.e. coming back to 
the same point of the sky on different timescales) and some prior 
assumptions about the slow drifts.  They are not efficient in removing 
fluctuations of the systematic effect on timescales smaller than the 
typical ``redundancy timescale", unless they have built in some filter 
which effectively does the equivalent of a high-pass filter on the 
timelines.

For a scan strategy such as that of Planck, a timeline-processing 
destriping technique as that of Cottingham 
\shortcite{cottingham-thesis} and Delabrouille 
\shortcite{jd-destripe98}, which would non-destructively reproject 
``near-optimally" each single timeline onto a map of the sky while 
removing slow drifts of various origin, may be quite unefficient in 
removing stripes in regions near the ecliptic plane, simply because of 
a lack of redundancy \cite{santanderjd}.  The Planck scanning indeed 
is such that each detector's field of view scans the sky on large 
($\sim 85$ degree radius) circles centered near the ecliptic.  The 
center of the circles is shifted by a few arcminutes every hour or so, 
the final field-of-view trajectory for each detector being a bundle of 
large circles intersecting each other near the ecliptic poles, with no 
intersection, or very few, in the region of the ecliptic (see figure 
\ref{fig:planck-scanning}).

In addition, removing correlated low-frequency drifts induced by the 
detector pickup of slow thermal fluctuations on the payload would be 
non-optimal, because it does not take advantage of the physical origin 
of the systematic which gives rise to correlations.  Taking care of 
this by using map making techniques as adapted versions of MADCAP 
\cite{madcap} which would treat all detectors simultaneously, taking 
into account noise (i.e. systematic effect) correlations is a 
possibility, but is very intensive computationnaly.  In this 
paper, we investigate the option which consists in treating the 
time-domain systematic effect as another component, ``living" in the 
time domain and not on two-D maps, taking care of the specificities 
this implies as compared to sky-attached, more traditionnal, 
astrophysical foregrounds.

\section[]{Simulations} \label{sec:sim}


\begin{figure*}
 \begin{center}
 \leavevmode
 \epsfxsize=\textwidth
 \epsfbox{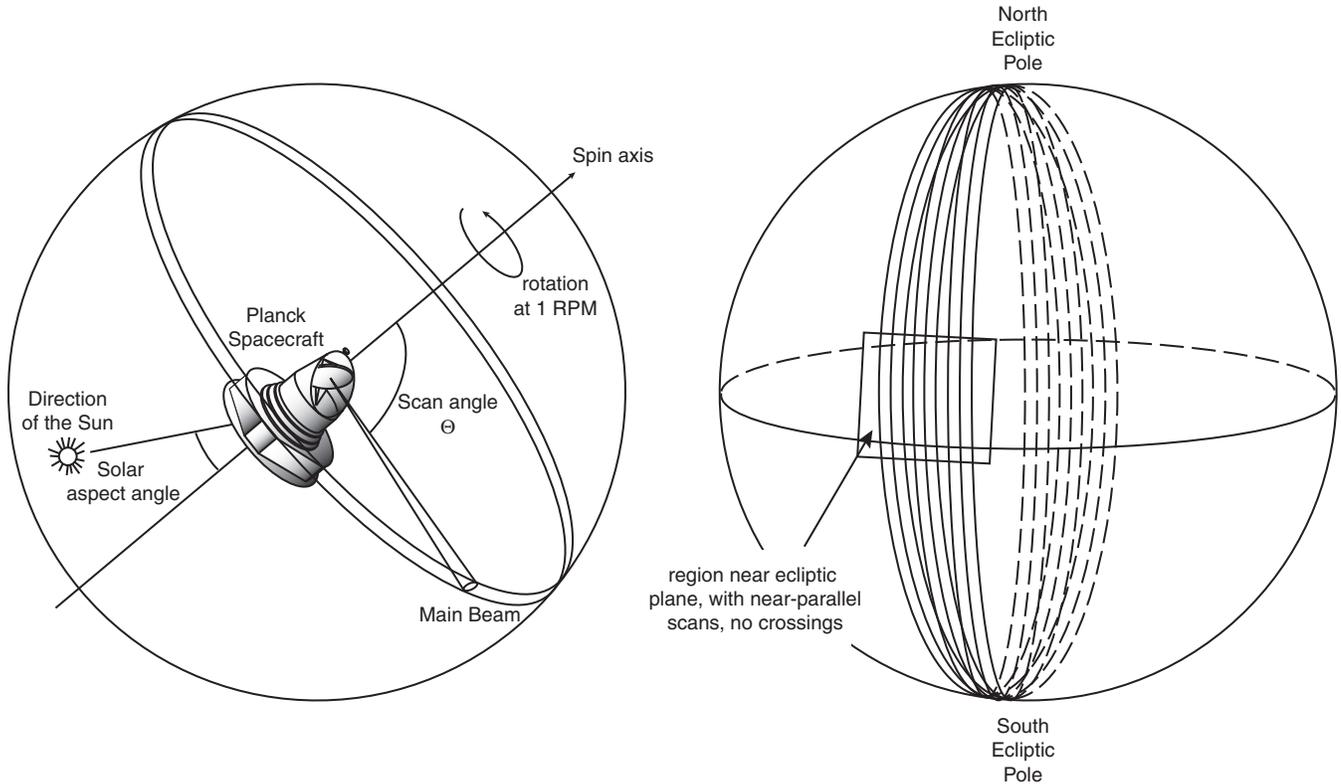}
 \end{center}
 \caption{Illustration of the Planck scanning strategy.  On the left 
 panel, the trajectory of the field of view of one single detector is 
 shown.  On the right panel, we show the trajectories of the centers 
 of the fields of views of three distinct detectors for three spin 
 axis position of Planck.  For adjacent spin axis positions, the 
 trajectories of all detectors are close to parallel for pointings 
 near the ecliptic plane.}
 \label{fig:planck-scanning}
\end{figure*}

We now investigate specifically the separation of instrumental and 
astrophysical components on simulations tailored for the Planck 
mission.  Much of the discussion can be straightforwardly adaptated to 
other experiments as Archeops or Boomerang.  The Planck mission, 
scheduled to be launched by ESA in 2007, will measure full-sky 
temperature and polarisation anisotropies of the sky emission in the 
frequency range 30 to 850 GHz.  The data will be taken using two 
instruments, the Low Frequency Instrument (LFI), and the High 
Frequency Instrument (HFI).  The LFI consists of 56 radiometers cooled 
to 20K, split into four frequency channels in the 30-100 GHz range 
\cite{mandolesi00,bersanelli00}, and the HFI of 48 bolometers split 
into 6 channels between 100 and 850 GHz \cite{lamarre00}.  This large 
frequency coverage, associated to unprecedented detector sensitivity, 
will allow to measure microwave anisotropies with a final sensitivity 
of about 6 $\mu$K per 8 arcminute pixel, provided all systematic 
effects remain under control.

We will investigate a component separation method that could be used 
for this mission among others, restricting ourselves to two sky 
components (CMB and thermal dust emission), one low frequency 
time-domain systematic effect due to the pickup of thermal emission 
from a part of the instrument, and realistic noise, including both 
white noise (in the time domain) and low-frequency detector noise 
(with a knee frequency of 0.01 Hz, as expected for Planck HFI 
detectors).  We will process the observations of the 48 HFI detectors 
only (6 channels) to avoid overconstraining excessively the problem.  
The method can be generalised to include LFI detectors as well, and 
more astrophysical (e.g. free-free emission, synchrotron, 
Sunyaev-Zel'dovich (SZ) effect \cite{sz}, another dust component, 
zodiacal emission) or time-domain (e.g. temperature of reference load, 
thermal bath fluctuations) components.

We implement the method on small square maps located near the ecliptic 
plane, convenient for a Fourier processing of the data.  The 
motivation for the choice of location is that in regions near the 
ecliptic poles, redundancies in the measurements are such that 
standard ``optimal" map making or destriping methods may be sufficient 
to remove the systematic effect, whereas we do not anticipate that 
this will be the case near the ecliptic.  We pick a region located at 
coordinates $\alpha = 204$ degrees and $\delta = 11$ degrees ($\lambda 
= 198$ degrees and $\beta = 19.5$ degrees), close to the ecliptic 
plane (for the scans to be almost parallel) and at a high galactic 
latitude (70 degrees).  We choose, for the specific illustration of 
our discussion here, a $272 \times 272$ pixels square map with two 
sides parallel to the equator, and a pixel size of 2.5 arcminutes.

The scanning strategy we adopt for generating our data is close to one 
of the possible Planck scanning strategies.  It consists in spinning 
the spacecraft at 1 RPM so that the beam axis, for each detector, 
scans the sky along large circles with 80 to 90 degrees angular radius 
(depending on the detector).  The spin axis is held fixed for about an 
hour, so that each of the detector main beams scans the same circle on 
the sky about 60 times.  Every hour, the spin axis is displaced by 
about 2.5 arcminutes to follow the apparent yearly motion of the Sun, 
and get a complete sky coverage in about 6 months.  Figure 
\ref{fig:planck-scanning} illustrates the Planck scanning strategy.  
Due to the detector setup in the focal plane, the scan-angle $\theta$ 
is slightly different for different detectors.

We now turn to generating synthetic observations of this region for 
all Planck HFI detectors, featuring CMB, dust, an instrumental 
effect, and detector noise.  

\subsection{Astrophysical components}


The CMB component is a 
COBE-normalised, randomly generated realisation of CMB anisotropies 
obtained using $C_{\ell}$ predicted by the CMBFAST software 
\cite{cmbfast} for a Universe with $H_{0}=65$ km/s/Mpc, 
$\Omega_{m}=0.3$, $\Omega_{b}=0.045$, $\Lambda=0.7$.  The CMB anisotropy 
emission is scaled as a function of frequency using the temperature 
derivative of the Planck law at T=2.726 K. 

The dust component is simply the extrapolation of the IRAS observed 
100 micron map (bright stars removed) with an assumed emission 
proportional to $\nu^2 \times B_{\nu}(T=17.5 K)$.  The sum of the two 
for each frequency channel gives the sky emission observable in that 
band in our simulation.  

Input templates for CMB and dust emission can be seen in figure
\ref{fig:results}.

We observe the simulated maps assuming locally a parallel scanning 
strategy such that the scans are distant, for a given detector, of 2.5 
arcminutes from each other.  We adopt the focal plane geometry 
described in \cite{lamarre00}.  For each detector, a purely 
astrophysical timeline is generated for each Planck detector by 
observing the properly weighted superposition of CMB and dust maps in 
a set of samples obtained each by integrating the sky emission in a 
symmetric Gaussian beam, properly positioned with respect to the map 
according to our selected scanning strategy. The beam size ranges from 
10 arcminutes (at 100 GHz) to 5 arcminutes (at 857 GHz). 

After this operation, we get a timeline in which only samples 
corresponding to observations in our generated map are non zero.  At 
equal times, the 48 detectors do not point towards the same point on 
the sky because of the focal plane geometry.  Shifts up to a few 
degrees, due to the angular extent of the image of the focal plane of 
the sky, separate the fields of views (FOVs) of the different 
detectors on the sky.

%

\subsection{Systematic effect}
A simulated instrumental foreground is generated in the form of a 
timeline corresponding to a plausible temperature fluctuation of an 
optical element: we generate a random, correlated gaussian stream, 
with a spectrum (as a function of frequency $f$) proportional to 
$f^{-2}$.  This is quite typical of temperature fluctuations observed 
in other experiments.  We assume that a prior timeline treatment as, 
e.g., that of Delabrouille \shortcite{jd-destripe98} has permitted the 
removal of the effect of such fluctuations on timescales larger than 
the spinning frequency of Planck, corresponding to one revolution of 
the satellite (one minute of time for a 1 RPM scanning).  In the end, 
the simulated instrumental effect due to temperature fluctuations has 
a $1/f^2$ spectrum between $f_{\rm spin} \simeq 0.017$ Hz and $f_{\rm 
sampling} \simeq 170$ Hz, and falls off to negligible values outside 
this interval.

The impact of the instrumental effect on each detector timeline 
depends on the value of a coupling coefficient for each detector.  
These coupling coefficients tells us how much the physical temperature 
fluctuations impact each detector reading.  They depend on the 
emissivity properties of the object whose temperature fluctuates (as a 
function of frequency $\nu$), and on the optical and geometrical 
properties of the setup.  We assume here a ${\nu}^{5/2}$ dependence 
(close to the Rayleigh-Jeans limit of a greybody with emissivity 
proportionnal to $\sqrt{\nu}$ as expected typically in our frequency 
range for aluminium).  The requirement on the temperature stability of 
the Planck payload, including the mirrors, is 0.2 mK. Such a 
temperature stability can be achieved in the very quiet space 
environment of the L2 Lagrange-point orbit of Planck, but not very 
easily for balloon-borne experiments as Archeops or Boomerang.  For 
our simulation, we assume a pessimistic temperature fluctuation of 
about 1 mK RMS integrated on timescales between 0.017 Hz and 170 Hz (5 
times higher than the expected upper limit for Planck), with the above 
$1/f^2$ spectrum and with a 3 \% total emissivity at 200 GHz.

This assumption of a strong instrumental effect, above expectations 
for Planck, make the dust contribution sub-dominant in all frequency 
channels.  This allows us to test the separation of a ``weak" 
astrophysical component.

Note that no detector dependence of the coupling within a frequency 
band is assumed in our simulation.  Such a dependence may exist in 
other cases as temperature fluctuations of a shield, for instance, 
where geometrical considerations have to be taken into account in the 
couplings.


\subsection{Detector noise}
For each detector, we generate a Gaussian coloured noise stream with a 
spectrum of the form $S_{n}(f) = K \times (1 + f_{\rm knee}/f)$, i.e. 
a typical detector noise with a white part and a $1/f$ part.  The knee 
frequency (frequency at which the two parts contribute equally) is 
taken to be $f_{\rm knee} = 0.01$ Hz, which corresponds to laboratory 
measurements performed for the Planck HFI. The noise is assumed to be 
uncorrelated between detectors.  Noise figures for the white noise 
level are representative of predictions for Planck HFI detectors.  
Again, we assume that very slow fluctuations have been removed, so 
that there remains no power outside the 0.017-170 Hz frequency band.  
One noise timeline is generated for each detector, and added 
to the appropriate data streams, which contain thus each the 
contributions from two astrophysical components, one instrumental 
component, and coloured noise.

\subsection{Reprojection on sky maps}
 For convenience, we will perform our treatment on two-dimensional 
 data rather than timelines.  In the present 
 implementation of the component separation, this allows to work 
 easily in the Fourier space while taking into account the different 
 detector pointings, in small regions of parallel scanning.
 We reconstruct a two-dimensional ``observed" map 
 for each detector by fitting to the observed timelines along with 
 corresponding pointings a quintic surface defined in each pixel, 
 matching best the observed data.  A map of the systematic effect 
 alone, obtained in the same way from the systematic effect timeline 
 of a fictitious detector centered on the optical axis of Planck, is 
 shown in figure \ref{fig:results}.

\begin{figure*}
 \begin{center}
 \leavevmode
 \epsfxsize=\textwidth
 \epsfbox{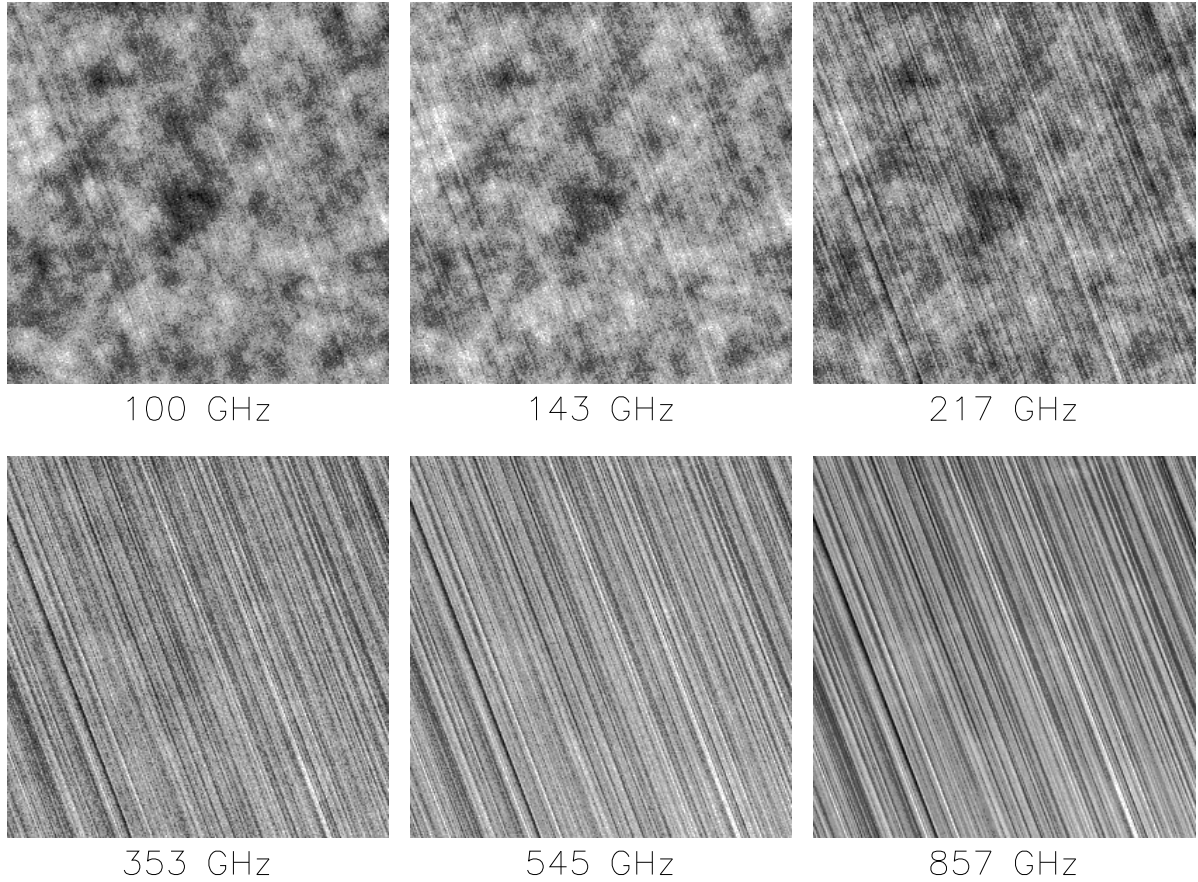}
 \end{center}
 \caption{Observations for one detector at each wavelength. For 
 illustrative purposes, we have chosen here a strong systematic 
 effect, which is comparable to CMB emission at low frequencies, and 
 dominates over dust at high frequencies.}
 \label{fig:obs-all-nu}
\end{figure*}

\begin{figure*}
 \begin{center}
 \leavevmode
 \epsfxsize=\textwidth
 \epsfbox{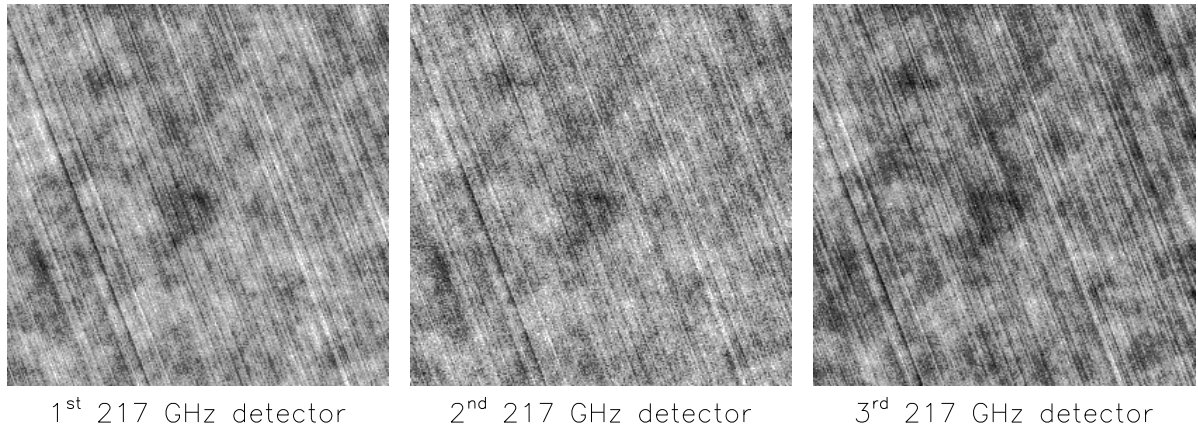}
\end{center}
\caption{Observations by three different detectors of the 217 GHz 
 channel.  In our example here, the observations at this frequency are 
 dominated by CMB anisotropies and stripes induced by the instrumental 
 effect .  The detector corresponding to the middle panel is polarised 
 and thus slightly noisier after recalibration of the observations, 
 because half of the input power (here assumed to be unpolarised) is 
 rejected.  The increased granularity due to high frequency noise can 
 (barely) be seen on the observation.  The detector-dependent shift of 
 the systematic effect with respect to sky emission is clearly visible 
 on these observed maps (look at the distance of the darkest stripe to 
 the upper left corner of the maps).}
 \label{fig:obs-one-nu}
\end{figure*}

In figure \ref{fig:obs-all-nu}, we display the reconstructed observed 
maps for one single detector of each channel.  These maps include the 
reprojected noise and systematic effect for each detector, as well as 
the observed astrophysical components.  In figure 
\ref{fig:obs-one-nu}, we display the observations for 3 detectors of 
the 217 GHz channel, which show the detector dependence of the 
observed systematic effect which, being synchronous on the timelines, 
is not coincident on the maps because of the shifts between detectors.

In table \ref{tab:StoN-all-channels}, we give for all channels the 
ratio between the RMS of the signal (CMB or dust) and the error 
performed on its measurement with our simulation setup.  The error 
contains contributions from detector noise, from the other 
astrophysical component, from the instrumental effect, and from 
observational errors (beam smoothing and reprojection effects).

\begin{table*}
 \begin{minipage}{150mm}
 \caption{Signal to Noise ratio for
 CMB and dust for observations in all channels. Here we consider as 
 ``noise" the error from all sources : detector noise, smoothing with 
 the beam, other components, and reprojection effects.}
 \label{tab:StoN-all-channels}
 \begin{center}
 \begin{tabular}{@{}lcccccccc}
 channel & 100 & 143 & 143p & 217 & 217p & 353 & 545p & 857\\
  \hline
  number of det. & 4 & 3 & 9 & 4 & 8 & 6 & 8 & 6 \\
  \hline
  $S/N$ CMB  & 1.4 & 1.2 & 0.95 & 0.77 & 0.65 & 0.16 & 0.01 & 
  $10^{-4}$ \\
  \hline
  $S/N$ dust  & $8.9 \times 10^{-3}$ & $2.0 \times 10^{-2}$ &  $1.7 \times 10^{-2}$ & 
  $5.9 \times 10^{-2}$ & $5.2 \times 10^{-2}$ & 0.14 & 0.22 & 0.26 \\
 \end{tabular}
 \end{center}
 \end{minipage}
\end{table*}

\section{Implementation of a component separation} \label{sec:imp-comp-sep}

\begin{figure}
    \begin{center}
    \leavevmode
    \epsfxsize=8.4cm
    \epsfbox{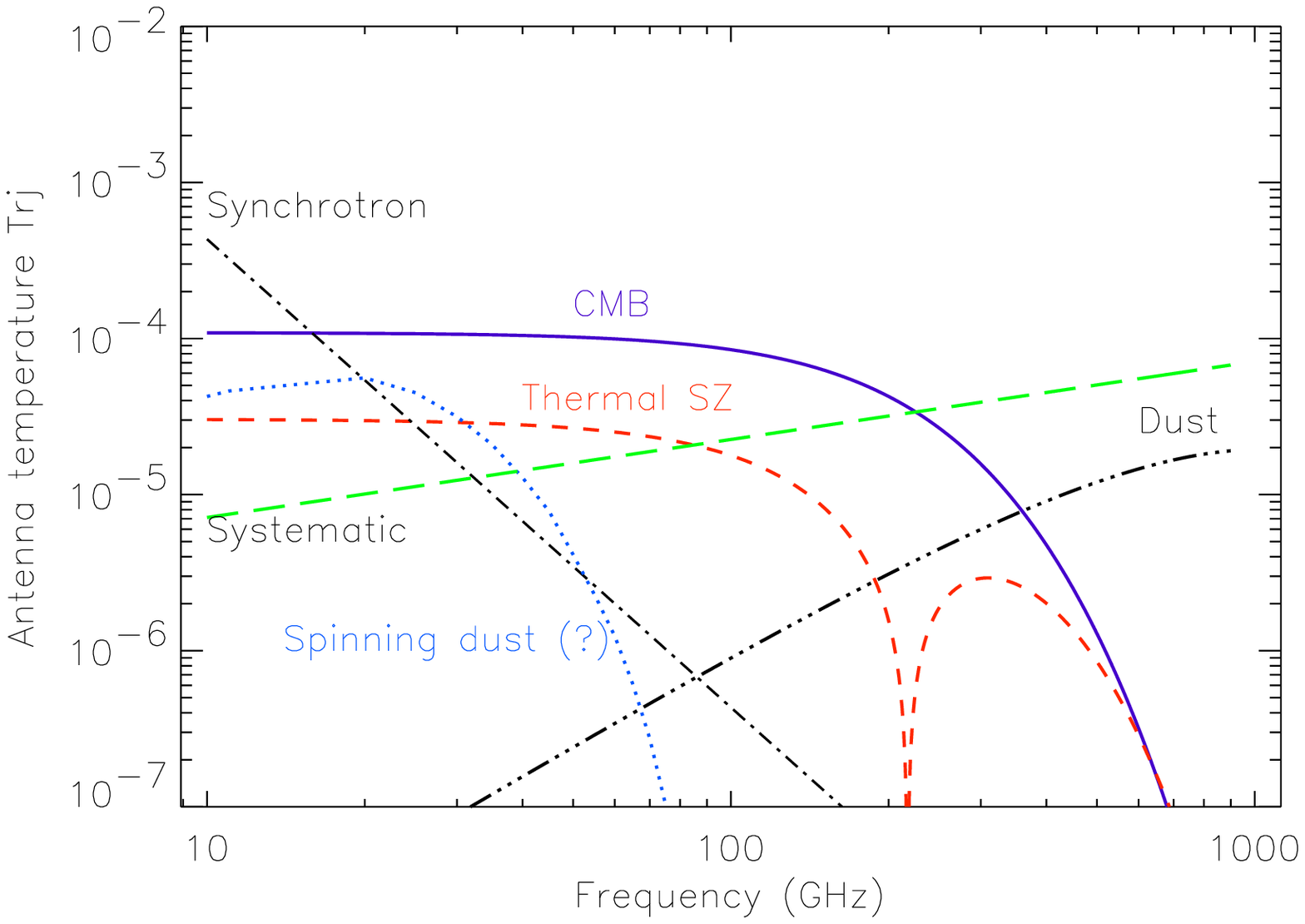}
    \end{center}
    \caption{Foreground emission in the frequency range of interest, 
    as compared to CMB anisotropy spectrum.  The existence of three 
    major galactic foregrounds, which are thermal dust emission, 
    synchrotron, and bremstrahlung (also called free-free, not 
    represented here), is well 
    established.  There is some evidence towards the existence of 
    microwave emission due to rotating dust.  SZ effect generates 
    foreground emission towards clusters of galaxies, and possibly 
    filamentary structure between them.  We have added a plausible 
    spectrum for a systematic effect due to temperature fluctuations of 
    an aluminium telescope mirror.  Relative levels are arbitrary, so 
    that the curve for each foreground emission can move up and down 
    depending on the level of contamination. The assumed shape of a 
    possible spinning dust component is from Draine \& Lazarian 
    \shortcite{drainelazarian}}
 \label{fig:foreground-spectra}
\end{figure}

 
The separation of astrophysical components we implement is possible 
essentially because the components have distinct emission spectra as a 
function of radiation wavelength.  This is illustrated in 
figure~\ref{fig:foreground-spectra}.  If the number of components and 
their respective spectra were perfectly well known, and if 
measurements were noiseless, multifrequency observations (with at 
least as many observations as there are components) would allow in 
principle the perfect recovery of the emission of each of the 
component at each point.  In practice, imperfections of the 
measurements and the model make the problem slightly more complicated.

\subsection{First principles}
In the case where astrophysical components only are present, we are 
given a number $N_{\nu}$ of observations of the same sky (or fraction 
of sky) at different frequencies $\nu$, each of which can be written as a 
linear superposition of emissions from the various astrophysical 
sources of radiation (CMB, dust, synchrotron, SZ\ldots) with some 
noise added:
\begin{equation}
    m_{\nu}({\vec {x}}) = A_{\nu c}T_{c}({\vec {x}}) + n_{\nu}({\vec {x}})
    \label{eq:simple-model-perfreq}
\end{equation}
This equation can equivalently be written in Fourier space, as
\begin{equation}
    m_{\nu}({\vec {k}}) = A_{\nu c}T_{c}({\vec {k}}) + n_{\nu}({\vec {k}})
    \label{eq:simple-model-perfreq-fourier}
\end{equation}
$A_{\nu c}$ is the ``mixing matrix" of this component separation 
problem, which is determined by the spectrum of the emission of the 
astrophysical components (up to a global calibration factor for each 
component) and by the optical response of the instrument.

If there are more observing frequencies than astrophysical components, 
and if the distributions for $n$ and $T$ are Gaussian and their 
autocorrelations $N$ and $S$ are known, the Wiener solution can 
be written as:
\begin{equation}
    {\widetilde T} = [S^{-1} + A^tN^{-1}A]^{-1} A^tN^{-1} \, m \equiv Wm
    \label{eq:wiener}
\end{equation}
Equation \ref{eq:wiener} defines the Wiener matrix 
$W = [S^{-1} + A^tN^{-1}A]^{-1} A^tN^{-1}$, which gives the 
best estimate ${\widetilde T}$ of the sky temperature $T$.

\subsection{Including instrumental effects}

Let us now assume that a number of instrumental effects are to be 
added to our model.  The measurement for each detector can be written 
as:

\begin{equation}
    m_{d}(\vec{x}) = A_{d c}T_{c}(\vec{x}) + A'_{d z}S_{z}(\vec{x},d) + n_{d}(\vec{x})
    \label{eq:complete-model-perdet2}
\end{equation}
where now $d$ indexes detectors.  $S_{z}(\vec{x},d)$ is the detector-dependent 
projection of the instrumental component. This ``reprojection" of the 
time-dependent instrumental component is not the same for all 
detectors because although all detectors experience the same 
instrumental effect (up to detector-dependent coupling factors) at the 
same time, they do not see the same pixel at the same time. Therefore, 
the systematic effect which pollutes the measured map in $\vec{x}$ is not the 
same for two different detectors $d$ and $d'$. Because of this, 
it is not possible to recombine simply the measurements of all the 
detectors at a given frequency into one single map.

In all generality, $S_{z}(\vec{x},d)$ depends on the scanning strategy 
and on the relative orientation of the focal plane and the scanning 
direction.  Therefore the relationship between reprojected effects 
$S_{z}(\vec{x},d)$ and $S_{z}(\vec{x},d')$ between two different 
detectors $d$ and $d'$ may be far from trivial (see appendix).  
However, if we restrict ourself to region of the sky where all the 
scans are parallel, there is a constant shift $\Delta \vec{x}_{d}$ 
between the direction pointed by detector $d$ and that of the optical 
axis, which is set entirely by the geometry of the image of the focal 
plane on the sky.  In that case equation 
(\ref{eq:complete-model-perdet2}) can be rewriten:

\begin{equation}
    m_{d}(\vec{x}) = A_{d c}T_{c}(\vec{x}) + A'_{d z}S_{z}(\vec{x}-\Delta \vec{x}_{d} ) + n_{d}(\vec{x})
    \label{eq:complete-model-perdet}
\end{equation}
where $S_{z}(\vec{x})$ is now the instrumental noise projected on the 
sky in the direction of the telescope optical axis.  Although it is 
not possible to average in bands without loosing information on the 
systematic, the system has a linear structure close to that 
of equation \ref{eq:simple-model-perfreq}, except for the additionnal 
systematic effect.

Rewriting equation \ref{eq:complete-model-perdet} in Fourier space, we 
get
\begin{equation}
    m_{d}({\vec {k}}) = A_{d c}T_{c}({\vec {k}}) + 
    A'_{d z} \exp(i \vec{k}.\vec{\Delta x_{d}}) S_{z}({\vec {k}}) + n_{d}({\vec {k}})
    \label{eq:complete-model-perdet-fourier}
\end{equation}
This has exactly the same structure as that of equation 
\ref{eq:simple-model-perfreq-fourier} if we include the spatial 
frequency dependent term $\exp(i \vec{k}.\vec{\Delta x_{d}})$ into 
matrix $A'$, and regroup matrices $A$ and $A'$ into one single matrix 
(and $(T_{c},S_{z})$ into one single vector). Therefore, we can apply 
a Wiener method to our extended problem. There are a few technical differences 
with the implementation of Bouchet and Gispert \shortcite{fb-rg99}:
\begin{itemize}
    \item in the present case, because of the phase shift for the 
    systematic component, the mixing matrix $A$ is complex;
    \item because of the problem asymmetry induced by
    the privileged scanning direction, the Wiener matrix $W$ 
    depends on $\vec{k}$ not simply on $|\vec{k}|$;
    \item the noise we assume is not white.
    \item we have one observed map for each detector, not one single 
    map per frequency band. 
    \item the observed maps are not simply the superposition of the 
    true pixellised maps with white noise added, but are 
    reconstructed from timelines and pointings by a global fit to the 
    data, which takes into account some of the inaccuracies of the
    map-making.
\end{itemize}

\subsection{Finite beam effect}
The effect of symmetric finite beams is to convolve the maps with a 
symmetric kernel $B({\vec {x}})$.  For asymmetric beams, the 
integration of the sky in the beam is scan-direction dependent, so 
that the operation is not strictly a convolution, except in the case 
of local parallel scanning.  In Fourier space, the sky temperature 
$T_{c}$ is multiplied by the Fourier transform of the asymmetric beam 
pattern (properly oriented).  The effect of the beam can therefore be 
taken straightforwardly into account by putting its ${\vec 
{k}}$-dependent effect into matrix $A$ (but not matrix $A'$, as the 
systematic effect is not affected by the shape of the beam).  Each 
component $A_{dc}$ of matrix $A$ needs to be replaced in equation 
\ref{eq:complete-model-perdet-fourier} by the product $A_{dc} 
B_{d}({\vec {k}})$ (no summation), where $B_{d}({\vec {k}})$ is the 
Fourier transform of the beam map for detector $d$.  We include the 
effect of the symmetric Gaussian beams in the evaluation of the mixing 
matrix $A$ for the inversion in Fourier space.

\subsection{Implementation}

The implementation of the Wiener method in Fourier space requires the 
apodisation of the maps to avoid border effects.  We take an 
apodisation function which goes to zero over 15 pixels (out of 272) on 
each side of the map, following a sine function from $\pi/2$ to 0.  
This apodisation induces little smoothing on the spectrum in ${\vec 
{k}}$-space, and leaves the inner region of the map untouched.

The implementation of the Wiener solution also requires prior 
information of the spectrum in ${\vec {k}}$-space, $S({\vec {k}})$, for all 
components.  For the CMB, we assume the true underlying spectrum 
(which differs slightly from the actual spectrum of our realisation 
because of the sample variance).  For dust, we assume a $1/(|{\vec 
{k}}|+C)^3$ law, where $C$ is a constant chosen to fit best the 
spectrum on the IRAS map.  For the systematic, we take the 
average of 10 realisations of a systematic according to the same 
generation law that we use for generating the actual observed 
systematic.  Therefore, for all components, we assume a spectrum which 
is close to the reality, but not exact (either because of sample 
variance, or because of different realisation, or a prior empirical 
fit).  We consider that this is representative of the  
uncertainty we will have when dealing with actual measurements. We 
discuss in more details the effect of uncertainties in the model of 
the measurement (and in particular uncertainties in pointing, 
frequency scalings) in the next section.

The simulation and inversion we implemented takes a few minutes of 
running time for $300 \times 300$ 2.5 arcminute pixels maps, on a very 
modest workstation.

\subsection{Results}
%
\begin{figure*}
    \begin{center}
    \leavevmode
    \epsfxsize=\textwidth
    \epsfbox{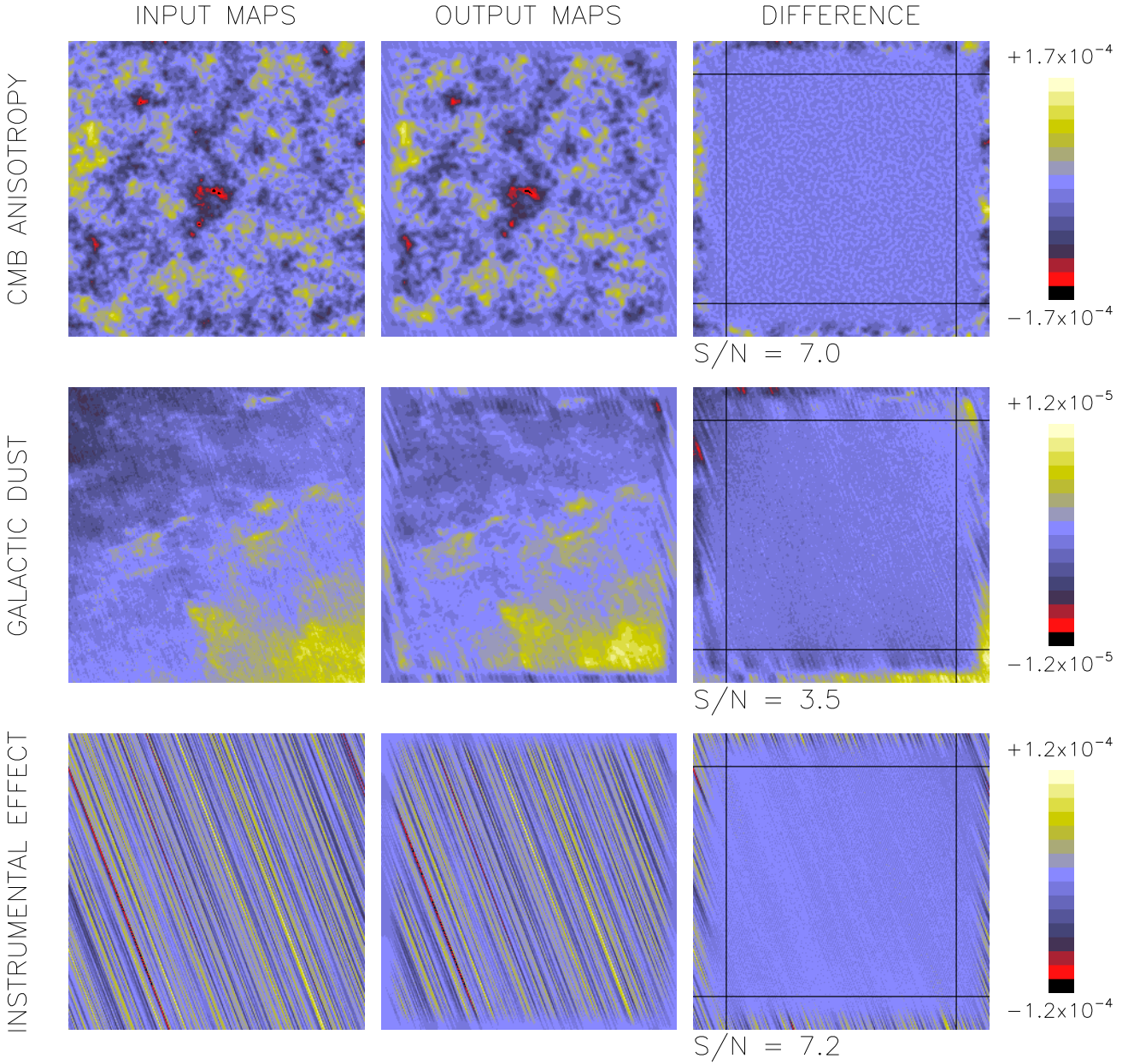}
    \end{center}
    \caption{Results of the inversion. The input maps are displayed in 
    the first column. Maps for CMB and dust are original maps, 
    whereas the map for the instrumental effect is a reprojection of 
    the effect's timeline. The output of the inversion, to be 
    compared with the original maps, are displayed in the second column. 
    The difference is shown in column three. For each component (on 
    each line), the three maps are displayed with the same greyscale 
    table (shown on the right of the figure) where figures give values 
    for 
    $\Delta{\rm T}/{\rm T}_{\rm CMB}$ at 200 GHz. Signal to noise 
    ratios of the final measurement in the center region (delimited by black 
    lines on the difference maps are given below the difference maps.)}
 \label{fig:results}
\end{figure*}

In figure \ref{fig:results}, we display the result of our inversion.  
Apart for border effects which here cannot be avoided, the original 
maps are well recovered.  Border effects are due both to the 
apodisation and to discontinuity effects of the instrumental effect 
map.  Apodisation effects appear only in a Fourier implementation, and 
could be avoided in real space implementations, at the cost of 
enormous required computation time.  The discontinuity effect on the 
other hand is not a problem of the implementation.  It appears because 
the map of the instrumental effect is not periodic, and the actual 
instrumental effect limited to the region of observation is thus not 
exactly the same for all detectors, because of detector shifts.  The 
effect remains small as long as detector shifts are small compared to 
the size of the map.

The RMS error on CMB anisotropies measurement, computed as the 
difference between the true input CMB maps and the output map 
(obtained from the reprojection of timelines and suffering from 
convolution with the beam, additive noise, and possibly residuals of 
other components after the separation process) is $\Delta T/T \simeq 
5.3 \times 10^{-6}$ in the central region of the map, corresponding to 
a signal to noise ratio of about 7.

Dust emission is very subdominant in all channels of our simulation, 
because we have chosen quite a strong instrumental effect.  This is 
unlikely to be the case for Planck, but may be typical of some 
balloon-borne experiments.  Despite of the fact dust can barely be 
seen in individual detector observations, its emission, in the central 
region, is recovered with a signal to noise ratio of about 3.5, and a 
$\Delta T/T$ of $5.3 \times 10^{-7}$ (at 200 GHz).

For both components, final results are much better than what could be 
expected from simple weighted averaging of all measured maps, as is clear 
when comparing the final $S/N$ rates with figures from table 
\ref{tab:StoN-all-channels}, and output maps (fig. \ref{fig:results}
with maps of the 
observations \ref{fig:obs-all-nu}.

The stripes due to the instrumental effect are well recovered, with a 
signal to noise of 7.2 in the central region of the map.  Of course, 
the sides of the map are not recovered as well, because of both the 
apodisation and the non-toric geometry of our maps, which invalidates 
the Fourier phase shift approximation of the actual dispacement of 
each detector at the borders.  

In table \ref{tab:StoN} we show the signal to noise ratio obtained at 
the end of the inversion, both for the 48-detector inversion and for a 
6-detector inversion with one detector per channel only.  For each 
case, we give $S/N$ computed in two ways.  The first way is to compute 
the error from the difference between the ``true" input map and the 
recovered map.  The second is to compute the error from the difference 
between the input map smoothed with the effective beam (see 
\ref{sec:beams}), and the recovered map.  Values obtained with the 
second method are given in parentheses.  The difference between the 
two is due to the part of the true astrophysical signal which is not 
recovered because of the imperfect resolution.

\begin{table}
 \begin{center}
 \caption{Final Signal to Noise ratio for
 CMB and dust, for both the 48-detector and the 6-detector inversions. 
 Figures in parentheses are obtained when the output is compared with 
 the original map smoothed with the effective beam.}
 \label{tab:StoN}
 \begin{tabular}{@{}lcc}
 component & CMB & dust\\
  \hline
  48-detector separation  & 7.0 (10.1) & 3.5 (5.1) \\
  \hline
  6-detector separation  & 4.1 (5.5) & 2.5 (3.7) \\
 \end{tabular}
 \end{center}
\end{table}

The difference between the signal to noise ratio for the 6-detector 
inversion and the 48-detector inversion is compatible with the loss of 
sensitivity due to the reduction of the number of detectors by a 
factor of 8.

In summary, the separation algorithm we implemented performs very 
satisfactorily, and can be a useful tool to constrain instrumental 
effects and separate various components in CMB data for a scanning 
experiment such as Planck.

\subsection{Effective beams} \label{sec:beams}
As $\widetilde{T} = Wm$, we have $\widetilde{T} = WAT + (\rm noise \; 
term)$, $W$ being the Wiener matrix and $A$ the mixing matrix.  This 
permits to compute an ``effective beam" for each component, which is 
the backward Fourier transform of the appropriate diagonal term of 
$W({\vec {k}})A({\vec {k}})$.

The final effective beam is due to a combination of the instrumental 
beam (which gets into matrix $A$) and of the attenuation of some 
spatial frequencies, where noise dominates, by the Wiener filter.

We compute effective beams for CMB and dust for both the 48-detector 
and the 6-detector inversions.  These effective beams are well fitted by 
two-dimensional elliptical Gaussians.  Because of the stripes, which 
increase the effective noise term on high frequencies perpendicular to 
the scanning, but not parallel to the scanning, beams are slightly 
elongated perpendicularly to the scanning, but this is not significant 
for the CMB. It is quite significant for the dust (especially in the 
case of the six-detectors inversion). This was to be expected, since 
the dust emission has a frequency dependence close to that of the systematic 
effect and a much lower amplitude.  Effective beam width for both 
inversions in co-scan and cross-scan directions are shown in table 
\ref{tab:beams}.

\begin{table}
 \begin{center}
 \caption{Effective co-scan and cross-scan beam sizes for 
 CMB and dust, for both the 48-detector and the 6-detector inversions.}
 \label{tab:beams}
 \begin{tabular}{@{}lcc}
 Direction & co-scan & cross-scan\\
  \hline
  48-det. CMB effective beam & 7.99 arcmin & 7.99 arcmin\\
  48-det. dust effective beam & 6.83 arcmin & 7.07 arcmin\\
  \hline
  6-det. CMB effective beam & 10.0 arcmin & 10.0 arcmin\\
  6-det. dust effective beam & 9.82 arcmin & 12.9 arcmin\\
 \end{tabular}
 \end{center}
\end{table}

The effective resolution depends on the number of detectors because in 
the case of 6 detectors, high frequency noise starts to dominate the 
signal at lower $\vec{k}$ values than when 48 detectors are present, 
which leads to effective beam degradation because of the Wiener 
filtering, which weights modes with a $S/(S+N)$ factor.  Results shown 
in this table are pessimistic for Planck, and especially so for dust 
because the instrumental effect, which dominates at high frequencies 
where dust is the highest astrophysical contribution, is quite high as 
compared to what is expected.

\section{Discussion} \label{sec:disc}

Our simulations have shown that parallel stripes due to the simulated 
systematic effect can be separated very well from astrophysical 
emission in the particular case we studied here. We now discuss 
the reasons why it works and the dependence of the result on specific 
assumptions.

 
\subsection{Impact of the number of components to be separated}
The test of our method as described in this paper has been done in a 
particularly favorable situation as far as the number of components is 
concerned.  We have 6 channels and 48 detectors for 3 components only.  
The system is thus widely overconstrained.  There is, however, no 
fundamental problem to adding more astrophysical components, as 
already demonstrated by Gispert and Bouchet \cite{fb-rg99}.  Here, we 
simply demonstrate the possibility to treat instrumental effects as 
additionnal components, provided the inversion is made on a system 
encompassing all detectors, and not co-averaged frequency channels.

We have performed tests of the method with the same number of 
components (two astrophysical and one systematic) but with a very 
limited number of detectors and channels (as little as three frequency
channels and six detectors), with satisfactory results. 

\subsection{Impact of the spectral properties of the systematic effect}

The  spectral properties of the astrophysical  components and of
the noise is an important issue regarding the separation procedure.

Let us imagine, for instance, that the systematic effect has almost 
the same frequency dependence as dust emission, e.g., an emission 
proportionnal to $\nu^4$.  This may happen for atmospheric emission in 
some cases of ballon-borne experiments, or in the case of Planck if 
some element of the spacecraft, which temperature fluctuates and which 
is coupled optically to the detectors, has a greybody emission with 
$\nu^2$ emissivity.  In this case, the frequency dependence does not 
help separating the two, and if all detectors were looking at the same 
point of the sky at the same time, there would be a very strong 
degeneracy between the dust and the thermal emission.  The best we 
could do in that case would be to filter stripes out of dust maps only 
based on spatial properties or on statistics (loosing at the same 
time some high frequency in the information on dust emission, in 
direction perpendicular to the stripes).  However, if as in the method 
we propose, the information concerning the different pointing 
direction of each detector is used, this degeneracy can be lifted to 
some extent.

\subsection{Further developpements}

The results we have presented here used a symmetric beam.  However, it 
is straightforward to extend our procedure to an asymmetric beam.  As 
long as scans are parallel, the effect on the maps of an asymmetric 
beam can be written, mathematically, as a convolution.  The Fourier 
components of the sky emissions are multiplied by the Fourier 
transform of the beam (properly oriented with respect to the scanning 
direction), which modifies their spectra.  This can be taken into 
account easily in the formalism of the separation, where the priors 
on components depend on $\vec{k}$ and not simply on $|\vec{k}|$.

The generalisation of our method to non-parallel scans is more 
difficult.  Our component separation method as implemented here works 
only for parallel equidistant scans.  In principle, small deviations 
from this assumption are not critical, as long as the deviations 
remain small as compared to the beam sizes.  The potential problem when 
the scanning departs from this ideal case is not the distortion of the 
assumed spectra for the Wiener solution, but the accuracy of the model 
of a detector-dependent constant displacement which permits to write 
the system in the simple form of equation 
\ref{eq:complete-model-perdet-fourier}.  We are currently working on 
relaxing the parallel-scanning assumption.

A first possible generalisation concerns a portion of the sky where 
two sets of parallel scans cross each other (which is often the case 
in actual CMB experiment).  In this case, it is possible to adapt our 
formalism by adding two systematic effect maps (one for each scanning 
direction) instead of one.

\section{Conclusions}

In this paper, we have investigated the possibility to use Wiener 
filtering to separate astrophysical components and systematic effects 
producing slow drifts in scanning CMB anisotropy experiments as the 
Planck mission.  We have shown that the ``component separation" 
formalism applied so far only to astrophysical emissions can be 
generalised to take into account and separate instrumental effects as 
well.

The algorithm we implemented is especially useful to remove systematic 
effects of known spectrum when there is little redundancy in the 
measurement for each detector channel, but high correlation between 
the systematic effect impact in the different detector channels. 
Whereas we have studied here the impact of one single systematic 
effect, that of slow temperature drifts of an optical element of the 
Planck telescope, the approach can be generalised and adapted to treat 
more effects, as for example one or more components due to atmospheric 
emission.

One of the difficulties for the application of our method is to get a 
good evaluation of the elements of the ``mixing matrix", i.e. matrix 
$A$, for the systematic effects.  For Planck, an accurate modeling 
associated to numerical simulations may permit to constrain these 
elements quite well, which may not be the case for other 
experiments/missions.  Methods 
permitting to estimate simultaneously mixing matrix elements and the 
components themselves \cite{baccigalupi00,snoussi01,maino01} can 
have a natural application in the context we discussed here.

\appendix

\section[]{General formulation of the full separation problem}

In this appendix, we write a general formalism for the separation of 
both sky-domain (astrophysical) and time-domain (instrumental) 
components for a scanning experiment.

In a scanning experiment such as those currently used for CMB mapping, 
it is usual to write the data stream of a given detector $d$ as a 
function of time as:

\begin{equation}
    s_{t} = M_{tp}I_{p} + n_{t}
\end{equation}

Where $s_{t}$ is the data stream for that detector (and $t$ indexes 
time), $I_{p}$ the brightness of a pixellised sky map in pixel $p$, 
and $M_{tp}$ the pointing matrix, which tells us how much pixel $p$ 
contributes to the data stream at time $t$. A summation convention is 
assumed over the repeated index $p$. $n_{t}$ is a noise stream, 
representing random noise, which may in some cases be coloured.

From a component separation point of view, the sky brightness is the 
sum of several astrophysical components, $I_{p} = A_{c}T_{cp}$. Here, 
$A_{c}$ is a single column of the ``mixing matrix" of the component 
separation problem, corresponding to one single detector $d$. The full 
mixing matrix for all detectors is noted $A_{dc}$.

Putting now explicitely in the equation the detector dependence as an 
additionnal index $d$, and writing explicitely the summations to 
avoid confusions, we get:

\begin{equation}
    s_{dt} = \sum_{p} \sum_{c} M_{dtp}A_{dc}T_{cp} + n_{dt}
    \label{eq:full-formalism1}
\end{equation}

For a satellite experiment as Planck, the size of matrix $A_{dc}$ is 
about $100 \times 6$, and the size of matrix $M_{dtp}$ about 
$100 \times 10^9 \times 10^7$, making it impractical to handle all 
the problems at the same time. As long as the noise of all detectors 
are independent, it is convenient to reproject timelines onto maps 
(one per detector, or possibly one per frequency channel) and then 
separate astrophysical components.

Instrumental effects as the one we discuss in this paper can be added 
to the formalism, in which case formula \ref{eq:full-formalism1} 
becomes:

\begin{equation}
    s_{dt} = \sum_{p} \sum_{c} M_{dtp}A_{dc}T_{cp} +
    \sum_{i} K_{di}x_{it} + n_{dt}
    \label{eq:full-formalism2}
\end{equation}
In this formula, $M_{dtp}$ is the pointing matrix for detector $d$, 
$A_{dc}$ the astrophysical component mixing matrix (which depends on 
the detector $d$ essentially through the corresponding frequency 
channel), $T_{cp}$ the sky emission of component $c$ in pixel $p$.
$x_{it}$ are the instrumental components (temperature fluctuations of 
spacecraft elements, atmospheric emission\ldots) where $i$ indexes 
components and $t$ indexes time, and
$K_{di}$ the mixing matrix for these instrumental components, 
which depends on the physical couplings between the sources and the 
detectors. $n_{dt}$ is the noise term.

In principle, equation \ref{eq:full-formalism2} (just as well as 
equation \ref{eq:full-formalism1}) can be rewritten in the general form
$s = AT + n$. The form of equation \ref{eq:full-formalism2}, however, 
shows more clearly the physics underlying the model.

\label{lastpage}


\begin{thebibliography}{99}

\bibitem[\protect\citename{Baccigalupi et al. }2000]{baccigalupi00}
   Baccigalupi, C. et al., 2000, MNRAS, 318, 769

\bibitem[\protect\citename{Beno\^{\i}t et al. }2001]{archeops-inst}
   Beno\^{\i}t, A. et al., 2001, to be published in Astroparticle Physics

\bibitem[\protect\citename{Bersanelli et al. }2000]{bersanelli00}
   Bersanelli, M. \& Mandolesi, N., 2000, Astro. Lett. and Communications, 37, 
   171
   
\bibitem[\protect\citename{Borrill }1999]{madcap}
   Borrill, J., 1999, in L. Maiani, F. Melchiorri, N. Vittorio, eds., 
   Proc. AIP Conf. 476, 3K Cosmology, New York, AIP, p. 277

\bibitem[\protect\citename{Bouchet \& Gispert }1999]{fb-rg99}
   Bouchet, F.R., Gispert, R., 1999, NewA, 4, 443

\bibitem[\protect\citename{Bouchet, Prunet \& Sethi }1999]{bps99}
   Bouchet, F.R., Prunet, S. and Sethi, Shiv K., 1999, MNRAS, 302, 663


\bibitem[\protect\citename{Cottingham }1987]{cottingham-thesis}
   Cottingham, D., 1987, PhD thesis, Princeton University, Princeton, New 
   Jersey

\bibitem[\protect\citename{de Bernardis et al. }1999]{boomerang-inst}
   de Bernardis, P., et al., 1999, New Astronomy Reviews, 43, 289

\bibitem[\protect\citename{Delabrouille }1998]{jd-destripe98}
   Delabrouille, J., 1998, A\&A Suppl. Ser., 127, 555
   
\bibitem[\protect\citename{Delabrouille, Gispert \& Puget }1998]{jd-moriond2}
   Delabrouille, J., Gispert, R., Puget, J.-L. in J. Tran Than Van, 
   Y. Giraud-H\'eraud, F. Bouchet, T. Damour, Y. Mellier,
   Proc. XXXIIIrd Rencontres de Moriond, Fundamental 
   parameters in Cosmology. Editions Fronti\`eres, p. 261

\bibitem[\protect\citename{Delabrouille et al. }2000]{santanderjd}
   Delabrouille, J. et al., 2000, Astro. Lett. and Communications, 37, 259

\bibitem[\protect\citename{Draine \& Lazarian }1998]{drainelazarian}
   Draine, B.T. \& Lazarian, A., 1998, ApJ letter, 494, L19

\bibitem[\protect\citename{Ferreira \& Jaffe }2000]{ferreira-jaffe00}
   Ferreira, P.G. and Jaffe, A.H., 2000, MNRAS, 312, 89

\bibitem[\protect\citename{Hanany et al. }2000]{Maxima}
   Hanany, S. et al.,  2000, ApJ letter, 545, L5 
   
\bibitem[\protect\citename{Hobson et al. }1998]{hjlb98}
   Hobson, M.P., et al., 1998, MNRAS, 300, 1
   
\bibitem[\protect\citename{Hu \& Sugiyama }1996]{hu-sugiyama96}
   Hu, W. and Sugiyama, N., 1996, ApJ, 471, 542

\bibitem[\protect\citename{Jungman et al. }1996]{jungman96}
   Jungman, G., et al., 1996,
   Phys. Rev. D, 54, 1332

\bibitem[\protect\citename{Lamarre et al. }2000]{lamarre00}
   Lamarre, J.-M. et al., 2000, Astro. Lett. and Communications, 37, 161

\bibitem[\protect\citename{Maino et al. }2001]{maino01}
   Maino, D. et al., 2001, submitted to MNRAS (astro-ph/0108362)

\bibitem[\protect\citename{Mandolesi et al. }2000]{mandolesi00}
   Mandolesi, N. et al., 2000, Astro. Lett. and Communications, 37, 151
   
\bibitem[\protect\citename{Schlegel, Finkbeiner \& Davies }1998]{SFD98}
   Schlegel, D. J., Finkbeiner, D. P. and  Davies, M., 1998, ApJ, 
   500, 525

\bibitem[\protect\citename{Seljak \& Zaldarriaga }2000]{cmbfast}
   Seljak, U., Zaldarriaga, M. , 2000, ApJ Suppl. ser.,  129, 431

\bibitem[\protect\citename{Silverberg }2000]{Tophat}
   Silverberg, R. F., 2000, Advances in Space Research, 26, 1401 

\bibitem[\protect\citename{Snoussi et al. }2001]{snoussi01}
   Snoussi, H. et al., 2001, to appear in MaxEnt 2001 proceedings
   (astro-ph/0109123)

\bibitem[\protect\citename{Sunyaev \& Zel'dovich }1972]{sz}
   Sunyaev, R., Zel'dovich, Y. , 1972, Comments Astrophys. Space 
   Phys., 4, 173



\bibitem[\protect\citename{Tegmark \& Efstathiou }1996]{tegmark-efstathiou96}
   Tegmark, M. and Estathiou, G., 1996, MNRAS, 281, 1297

\bibitem[\protect\citename{Vielva et al. }2001]{vielva01}
   Vielva, P. et al., 2001, MNRAS, 326, 181

\bibitem[\protect\citename{Wright }1999]{Map}
   Wright, E. L., 1999, New Astronomy Reviews, 43, 257    

\end{thebibliography}
\end{document}